\documentclass{eptcs}

\providecommand{\event}{MSFP 2016} 

\usepackage[all]{xy}
\usepackage{stmaryrd}
\usepackage{amssymb}

\usepackage{proof}

\newcommand{\eqdf}{=_{\mathrm{df}}}

\newcommand{\Set}{\mathbf{Set}}
\newcommand{\Cat}{\mathbf{Cat}}

\newcommand{\Cont}{\mathbf{Cont}}
\newcommand{\DCont}{\mathbf{DCont}}
\newcommand{\Poly}{\mathbf{Poly}}
\newcommand{\DPoly}{\mathbf{DPoly}}

\newcommand{\Comonad}[1]{\mathbf{Comonads({#1})}}

\newcommand{\csem}[1]{\llbracket #1 \rrbracket^\mathrm{c}}
\newcommand{\dcsem}[1]{\llbracket #1 \rrbracket^\mathrm{dc}}

\newcommand{\ia}[1]{\{#1\}}

\newcommand{\C}{\mathbb{C}}
\newcommand{\D}{\mathbb{D}}
\newcommand{\J}{\mathbb{J}}

\newcommand{\id}{\mathsf{id}}

\newcommand{\op}{\mathrm{op}}

\newcommand{\inv}{{-1}}

\newcommand{\zt}{{\ast}}

\newcommand{\inl}{\mathsf{inl}}
\newcommand{\inr}{\mathsf{inr}}

\newcommand{\eps}{\varepsilon}
\newcommand{\de}{\delta}

\newcommand{\dn}{\downarrow}
\renewcommand{\o}{\mathsf{o}}
\newcommand{\pl}{\oplus}

\newcommand{\dnp}{\mathbin{\downarrow'}}
\newcommand{\plp}{\mathbin{\oplus'}}

\newcommand{\Pt}{\bar{P}}
\newcommand{\qt}{\bar{q}}

\newcommand{\s}{\mathsf{s}}
\renewcommand{\t}{\mathsf{t}}

\newcommand{\Nat}{\mathsf{Nat}}
\newcommand{\Str}{\mathsf{Str}}
\newcommand{\NEList}{\mathsf{NEList}}

\begin{document}

\title{Directed Containers as Categories}
\author{Danel Ahman
\institute{LFCS, School of Informatics, University of Edinburgh, United Kingdom}
\email{d.ahman@ed.ac.uk}
\and Tarmo Uustalu
\institute{Institute of Cybernetics at Tallinn University of Technology, Estonia}
\email{tarmo@cs.ioc.ee}
}
\def\titlerunning{Directed Containers as Categories}
\def\authorrunning{D.~Ahman \& T.~Uustalu}

\maketitle

\begin{abstract}
  Directed containers make explicit the additional structure of those
  containers whose set functor interpretation carries a comonad
  structure. The data and laws of a directed container resemble those
  of a monoid, while the data and laws of a directed container
  morphism those of a monoid morphism in the reverse direction. With
  some reorganization, a directed container is the same as a small
  category, but a directed container morphism is opcleavage-like. We
  draw some conclusions for comonads from this observation,
  considering in particular basic constructions and concepts like the
  opposite category and a groupoid.
\end{abstract}

\section{Introduction}

Abbott, Altenkirch, Ghani, Hancock, McBride and Morris's containers
\cite{AAG:conspt,AGHMM:indc} are a representation of a wide class of
set functors (datatypes) in terms of shapes and positions. We could
say that they are a form of syntax for datatypes as semantic
entities. Polynomials \`a la Gambino and Hyland \cite{GH:weltdp} are a
close variation of containers. They are nice in particular because
they can be made to work in a very general setting (categories with
pullbacks) \cite{Web:polcp}, but the definitions and proofs get very
involved very early on.

Ahman, Chapman and Uustalu \cite{ACU:wheicc} introduced directed
containers to make explicit the additional structure of those
containers whose set functor interpretation carries a comonad
structure. Typical examples of such functors with a comonad structure
are node-labelled tree datatypes. The counit extracts the root label
of a tree while the comultiplication relabels every node with the
whole subtree rooted by that node. In a directed container every
position in a shape determines another shape, the subshape
corresponding to this position. Ahman, Chapman and Uustalu
\cite{ACU:wheicc} also spelled out the polynomial version of directed
containers (directed polynomials), but did not put it to any use.

In this paper, we revisit directed polynomials and take advantage of
them to make some observations about the structure of the category of
directed containers. In particular, we develop some new constructions
on directed containers and some specializations of directed
containers.
Specifically, we see that the category of directed containers is
isomorphic to the category of small categories and (what we here call)
relative split pre-opcleavages of a certain type. We consider the
opposite category construction (for a datatype of node-(i.e.,
subtree-)labelled trees, the corresponding datatype of
context-labelled trees) and the concept of a groupoid (of which zipper
datatypes are typical examples).

The significance of containers for functional programming consists
primarily in providing a tool for generic programming with both datatypes 
in general as well as with datatypes with special structure. But containers 
are also particularly well-suited for analyzing special classes of datatypes 
from a combinatorial perspective. For instance, the concepts of directed
containers and distributive laws of directed containers enable us to
recognize and enumerate the comonad structures that are available on
different functors, the distributive laws that exist between these
comonads etc.

This paper is organized as follows. First, in Section~\ref{sect:dcontainers}, we briefly review
containers and directed containers. Then, we go to polynomials and
directed polynomials in Section~\ref{sect:dpolynomials}, recognizing that a directed polynomial is just a small category while a directed polynomial morphism is not a functor, but something different and considerably more involved. Next, in Section~\ref{sect:coproductandtensor}, we consider the coproduct of two directed
containers and a certain tensor of two directed containers as constructions of categories. Finally, we
analyze opposite categories and groupoids as directed containers in Sections~\ref{sect:oppositedcontainer} and~\ref{sect:bdcontainers}, respectively.


Throughout the paper, we use syntax similar to that of the
dependently-typed functional programming language Agda. In particular,
when declaring the type of a function, we mark the arguments that are
derivable in most contexts as implicit by enclosing them/their types
in braces. In applications of the function, we either omit these
arguments or provide them in braces. In infix applications of
functions, we place the implicit arguments (if we want to provde them)
after the function symbol.

\section{Directed Containers}
\label{sect:dcontainers}

We begin with a brief recap of containers and directed containers.

\subsection{Containers}

A \emph{container} is given by a set $S$ (of shapes) and a $S$-indexed
family of sets $P$ (of positions).
Any container $(S, P)$ interprets into a set functor $\csem{S, P}$,
defined on objects as $\csem{S, P}\, X \eqdf \Sigma s : S.\, P\, s \to X$ and
on morphisms as $\csem{S, P}\, f\, (s, v) \eqdf (s, \lambda p.\, f\, (v\, p))$.

A \emph{container morphism} between $(S, P)$ and $(S', P')$ is given
by maps $t : S \to S'$ and \linebreak $q : \Pi s : S.\, P'\, (t\, s) \to P\, s$, 
%
correspondingly interpreting 
into a natural transformation $\csem{t, q}$ between
the functors $\csem{S, P}$ and $\csem{S', P'}$, defined as 
$\csem{t, q}\, (s, v) \eqdf (t\,s, \lambda p.\, v\, (q\, s\, p))$.

Containers and container morphisms form a monoidal category
$\Cont$. The interpretation $\csem{-}$ is a fully faithful monoidal
functor from $\Cont$ to $[\Set,\Set]$, with the monoidal structures
given by the composition of containers and the composition of set
functors, respectively.

We emphasize the fully-faithfulness of $\csem{-}$: it means that all natural
transformations between functors representable by containers are
representable as container morphisms and uniquely so.

\subsection{Directed Containers}

Directed containers are containers with additional structure.  

A \emph{directed container} is a container $(S, P)$ together with maps
${\dn} : \Pi s : S.\, P\, s \to S$ (the subshape corresponding to a
position), $\o : \Pi \ia{s : S}.\, P\, s$ (the root position in a shape), ${\pl} : \Pi \ia{s
  : S}.\, \Pi p : P\, s.\, P\, (s \dn p) \to P\, s$ (translation of
subshape positions into the global shape) such that
\[
\begin{array}{c}
s \dn \o = s \\
s \dn (p \pl p') = (s \dn p) \dn p' \\
p \pl \o = p \\
\o \pl p = p \\
(p \pl p') \pl p'' = p \pl (p' \pl p'')
\end{array}
\]
where the 4th displayed equation is welltyped thanks to the 1st (to type the
left-hand side we need $p : P\, (s \dn \o)$ while to type the
right-hand side we need $p : P\, s$) and the 5th thanks to the 2nd (to
type the left-hand side we need $p'' : P\, (s \dn (p \pl p'))$ while
to type the right-hand side we need $p'' : P\, ((s \dn p) \dn p)$). In
the special case $S = 1$, the equations simply require that $(P\, \zt,
\o\, \ia{\zt}, {\pl}\, \ia{\zt})$ is a monoid.

Any directed container $(S, P, {\dn},\o, {\pl})$ interprets into a comonad $\dcsem{S, P, {\dn},
  \o, {\pl}} \eqdf (D, \eps, \de)$ on $\Set$, with $D \eqdf \csem{S, P}$,
$\eps\, (s, v) \eqdf v\, (\o\, \ia{s})$, $\de\, (s, v) \eqdf (s,
\lambda p.\, (s \dn p, \lambda p'.\, v\, (p \pl \ia{s}\, p')))$.

A \emph{directed container morphism} between $(S, P, \dn, \o, \pl)$
and $(S', P', {\dnp}, \o', {\plp})$ is a container morphism $(t, q)$
between $(S, P)$ and $(S', P')$ such that
\[
\begin{array}{c}
t\, (s \dn q\, s\, p) = t\, s \dnp p \\
\o\, \ia{s} = q\, s\, (\o'\, \ia{t\, s}) \\
q\, s\, p \pl \ia{s}\, q\, (s \dn q\, s\, p)\, p' = q\, s\, (p \plp \ia{t\, s}\, p')
\end{array}
\]
where the 3rd displayed equation is welltyped as the 1st holds (to type the 
left-hand side we need $p' : P'\, (t\, (s \dn q\, s\, p))$ while to type the 
right-hand side we need $p' : P'\, (t\, s \dnp p)$). In the special
case $S = S' = 1$, which trivializes $t$, the map $q\, \zt$ is a
monoid morphism between $(P'\, \zt, \o'\, \ia{\zt}, {\plp}\, \ia{\zt})$ and $(P\, \zt, \o\, \ia{\zt}, {\pl}\, \ia{\zt})$---in the opposite
direction compared to the directed container morphism.

A directed container morphism $(t,q)$ between $(S, P, \dn, \o, \pl)$
and $(S', P', {\dnp}, \o', {\plp})$
interprets into a comonad morphism
between $\dcsem{S, P, \dn, \o, \pl}$ and $\dcsem{S', P', {\dnp}, \o',
  {\plp}}$, given by $\dcsem{t, q} \eqdf \csem{t, q}$.

Directed containers and directed container morphisms form a category
$\DCont$. The interpretation
$\dcsem{-}$ is a fully faithful functor from $\DCont$ to
$\Comonad{\Set}$. In fact, $\dcsem{-}$ is a pullback of $\csem{-}$ along 
$U : \Comonad{\Set} \to [\Set,\Set]$, meaning that directed 
containers are precisely those containers whose set functor interpretation 
carries a comonad structure.

Here are some simple examples of directed containers.

Taking $S \eqdf 1$, $P\, \zt \eqdf \Nat$, $\zt \dn p \eqdf \zt$, $\o
\eqdf 0$, $p \pl p' \eqdf p + p'$, we get a directed container
interpreting into the stream comonad: $D\, X \eqdf \Sigma \zt : 1.\,
\Nat \to X \cong \Str\, X$. The counit extracts the head element from
a stream while the comultiplication turns a stream into a stream of
its suffixes.

Taking $S \eqdf \Nat$, $P\, s \eqdf [0..s]$, $s \dn p \eqdf s - p$,
$\o \eqdf 0$, $p \pl p' \eqdf p + p'$, we get a directed container
interpreting into the nonempty list comonad: $D\, X \eqdf \Sigma s :
\Nat.\, [0..s] \to X \cong \NEList\, X$. The counit extracts the head
element from a nonempty list while the comultiplication turns a
nonempty list into a nonempty list of its suffixes.

If we instead take $S \eqdf \Nat$, $P\, s \eqdf [0..s]$, $s \dn p \eqdf
s$, $\o \eqdf 0$, $p \pl \ia{s}\, p' \eqdf (p + p')
\mathbin{\mathrm{mod}} (s + 1)$, we get a directed container
interpreting into a different comonad on the nonempty list
functor. The counit extracts the head element from a nonempty list
while the comultiplication turns a nonempty list into a nonempty list
of its cyclic shifts.

Finally, taking $S$ to be any given set and $P\, s \eqdf 1$, we get a
directed container interpreting into the reader comonad $D\, X \eqdf
\Sigma s : S.\, 1 \to X \cong S \times X$. Taking $S$ to be any given
set, $P\, s \eqdf S$, $s \dn s' \eqdf s'$, $\o\, \ia{s} \eqdf s$ and
$s' \pl \ia{s}\, s'' \eqdf s''$, the directed container obtained
interprets into the array comonad $D\, X \eqdf \Sigma s : S.\, S \to X
\cong S \times (S \to X)$.

\section{Directed Containers as Categories}
\label{sect:dpolynomials}

We now proceed to polynomials and directed polynomials.

\subsection{Containers as Polynomials}

Polynomials are an alternative view of containers where the $S$-indexed 
family $P$ of positions is replaced with a single set
$\Pt$ that collects all positions across all shapes. This set must be
fibred over $S$ in the sense of coming together with a map $\s : \Pt
\to S$ that tells which shape each position belongs to.

A \emph{polynomial} is hence given by two sets $S$ and $\Pt$ and a map
$\s: \Pt \to S$.

Containers can be converted into polynomials and vice versa by $\Pt
\eqdf \Sigma s : S.\, P\, s$, $\s\, (s, p) \eqdf s$ resp.\ $P\, s
\eqdf \Sigma p : \Pt.\, \ia{\s\, p = s}$.\footnote{Elements of $P\, s$ are dependent pairs with an equality proof as the second component. We mark this component as implicit by enclosing its type in braces in the definition of $P\, s$ and omit it in actual pairs of this type. We apply the same practice also to other similar dependent tuple types.} This gives us a bijection up to
isomorphism between the collections of containers and polynomials.

In a morphism between polynomials, instead of an $S$-indexed family of
maps $q\, s : P'\ (t\, s) \to P\, s$, we have a single map $\qt$
sending elements of $\Pt'$ to $\Pt$ which is defined only for those
$p : \Pt'$ whose shape is in the image of $t : S \to S'$, i.e., $t\, s
= \s'\, p$ for some $s : S$. This $s$ must be explicitly supplied, as
it is not uniquely determined by $p$, unless $t$ is
injective. Moreover, the shape for the position returned by $\qt$ on
$(s, p)$ must be $s$, i.e., $\s\, (\qt\, (s, p)) = s$.\footnote{These
  considerations concern general polynomial morphisms, which is what
  we need. In works on polynomials, it is commonplace to consider only
  linear polynomial morphisms. They are much easier to handle, but too
  restrictive for our purposes.}

A \emph{polynomial morphism} between $(S, \Pt, \s)$ and $(S', \Pt',
\s')$ is therefore given by maps $t : S \to S'$ and $\qt : (\Sigma s :
S.\, \Sigma p : \Pt'.\, \ia{t\, s = \s'\, p}) \to \Pt$ such that $\s\,
(\qt\, (s, p)) = s$. Polynomials and polynomial morphisms form a
category $\Poly$.

Container morphisms and polynomial morphisms are interconverted by
$\qt\, (s, p) \eqdf q\, s\, p$ resp.\ $q\, s\, p \eqdf \qt\, (s,
p)$. These conversions extend the bijection up to isomorphism between
the collections of containers and polynomials to an equivalence
between the categories $\Cont$ and $\Poly$ of containers resp.\
polynomials.

\subsection{Directed Containers as Directed Polynomials as Small
  Categories}

Let us apply a similar change of view (from ``indexed'' to ``fibred'')
to directed containers. We arrive at the following result.

A \emph{directed polynomial} is given by sets $S$, $\Pt$ and maps
$\s, \t : \Pt \to S$, $\id : \ia{S} \to P$, ${;} : (\Sigma p : \Pt.\,
\Sigma p' : \Pt.\, \ia{\t\, p = \s\, p'}) \to \Pt$ such that $\s\, (\id\,
\ia{s}) = s$, $\s\, (p \mathbin{;} p') = \s\, p$ and 
\[
\begin{array}{c}
\t\, (\id\, \ia{s}) = s \\
\t\, (p \mathbin{;} p') = \t\, p' \\
p \mathbin{;} \id\, \ia{s} = p \\
\id\, \ia{s} \mathbin{;} p = p \\
(p \mathbin{;} p') \mathbin{;} p'' = p \mathbin{;} (p' \mathbin{;} p'')
\end{array}
\]
Similarly to the situation with directed containers, here the 4th
displayed equation is welltyped because the 1st equation holds and the
5th equation is welltyped because the 2nd equation holds. (But the
more basic nondisplayed equations need also to be used.)

A directed container is converted into a directed polynomial by $\Pt
\eqdf \Sigma\, s : S.\, P\, s$, $\s\, (s, p) \eqdf s$, $\t\, (s, p)
\eqdf s \dn p$, $\id\, \ia{s} \eqdf (s, \o\, \ia{s})$, $(s, p)
\mathbin{;} (s \dn p, p') \eqdf (s, p \pl \ia{s}\, p')$.

A conversion in the opposite direction is given by $P\, s \eqdf \Sigma
p : \Pt.\, \ia{\s\, p = s}$, $s \dn p \eqdf \t\, p$, $\o\, \ia{s} \eqdf
\id\, \ia{s}$, $p \pl \ia{s} p' \eqdf p \mathbin{;} p'$. These
conversions form a bijection up to isomorphism between the collections
of directed containers and directed polynomials.

It is easy to notice that the data and laws of a directed polynomial
are just those of a small category: shapes are objects, positions are
morphisms, $\s$ and $\t$ give the source and target of a morphism, $\id$ and
${;}$ are the identities and composition. We learn that the reason why
a directed container is like a monoid, albeit not quite, is that it is
a proper generalization of a monoid, and a well-known proper
generalization at that, a small category!

What are the implications of this? Can it be that the category of
directed containers is nothing but the category $\Cat$ of small
categories and functors?  That would be a hasty conclusion. We should
diligently spell out what a directed container morphism amounts to. It
is this.

A \emph{directed polynomial morphism} between $E = (S, \Pt, \s, \t,
\id, ;)$ and $E' = (S', \Pt', \s', \t', \id', ;')$ is given by
maps $t : S \to S'$ and $\qt : (\Sigma s : S.\, \Sigma p : \Pt'.\, \ia{t\,
s = \s'\, p)} \to \Pt$ such that $\s\, (\qt\, (s, p)) = s$ and 
\[
\begin{array}{c}
t\, (\t\, (\qt\, (s, p))) = \t'\, p \\
\id\, \ia{s} = \qt\, (s, \id'\, \ia{t\, s}) \\
\qt\, (s, p) \mathbin{;} \qt\, (\t\, (\qt (s, p)), p') = \qt\, (s, p \mathbin{;'} p')
\end{array}
\]
where the 3rd displayed equation is welltyped because the 1st holds.
Pictorially,
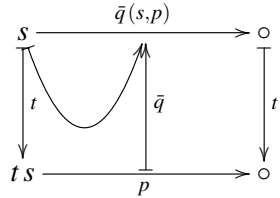
\[
\xymatrix@R=1.5pc@C=3pc{
s \ar@{->}[rr]^-{\qt\, (s, p)} \ar@{|->}[dd]^t \ar@{|->}@/_3pc/[r] 
    & & \circ \ar@{|->}[dd]^t\\
    & & \\
t\, s  \ar[rr]_{p} 
    &  \ar@{|->}[uu]_\qt & \circ
}
\]
Directed polynomials and directed polynomial morphisms form a
category $\DPoly$.\footnote{Linear directed polynomial morphisms are much
  simpler than general directed polynomial morphisms: they are nothing but
  fully-faithful functors between small categories.}

Directed container morphisms and directed polynomial morphisms are
interconvertible by $\qt\, (s, p) \eqdf q\, s\, p$ resp.\ $q\, s\, p
\eqdf \qt\, (s, p)$. This extends the bijection up to isomorphism
between the collections of directed containers and directed
polynomials into an equivalence of the categories $\DCont$ and
$\DPoly$.

We see that a directed polynomial morphism $(t, \qt)$ between directed
polynomials $E$ and $E'$ is very far from anything like a functor
between $E$ and $E'$ as small categories. Instead, it is a bit like
$\qt$ being an opcleavage for $t$, but with two big reservations.
First, $t$ is not a functor, but only an object mapping, and second,
the requirements on $\qt$ are somewhat weak.

We could reasonably say that $\qt$ is a split pre-opcleavage for
$t^\dagger: E \to S'^\dagger$ relative to $! : E' \to S'^\dagger$
where $S'^\dagger$ is the cofree category on $S'$, i.e., the small
category with $S'$ as the set of objects and a single morphism between any
two objects.  This is according to the following terminology that we
have invented for the occasion.

We say that a \emph{split pre-opcleavage} for a functor $F : \C \to
\D$ is an assignment, to any object $X$ of $\C$ and any morphism $f$ of
$\D$ with $F\, X$ as the source, of a morphism $f_*\, X$ of $\C$ that has
$X$ as the source and satisfies $F\, (f_*\, X) = f$, in such a way
that $\id\, \ia{X} = (\id\, \ia{F\, X})_*\, X$ and $f_*\, X ; g_*\, Y
= (f ; g)_*\, X$. This is weaker than an opcleavage in that $f_*\ X$
does not have to be opCartesian. But at the same time, this is also
stronger in that the equations of a split opcleavage are already
required. As a variation, a \emph{split pre-opcleavage} for a functor
$F: \C \to \D$ \emph{relative} to a functor $J : \J \to \D$ is an
assignment, to any object $X$ of $\C$ and any morphism $f$ of $\J$ whose
source $Z$ satisfies $F\, X = J\, Z$, of a morphism $f_*\, X$ of $\C$ that
has $X$ as the source and satisfies $F\, (f_*\, X) = J\, f$, in such a
way that $\id\, \ia{X} = (\id\, \ia{F\, X})_*\, X$ and $f_*\, X ;
g_*\, Y = (f ; g)_*\, X$. Here the meaning of `relative'
is the same as in relative adjuctions \cite{Ulm:prodra} and relative
monads \cite{ACU:monnnb}.

We are not sure at this stage that this is the optimal analysis of
directed polynomial morphisms in terms of standard or close to
standard concepts, but it is the best we currently
have.  

Let us revisit our examples. The small category for the stream comonad
has $S \eqdf 1$, $\Pt \eqdf \Nat$, $\s\, p \eqdf \zt$, $\t\, p \eqdf
\zt$, $\id \eqdf 0$, $p \mathbin{;} p' \eqdf p + p'$. (Of course this
is nothing else than the monoid $(\Nat, 0, +)$ seen as one-object
category.)


The small category for the nonempty lists and suffixes comonad is
given by $S \eqdf \Nat$, $\Pt \eqdf \Sigma s : \Nat.\, [0..s]$, $\s\,
(s, p) \eqdf s$, $\t\, (s, p) \eqdf s - p$, $\id\, \ia{s} \eqdf (s,
0)$, $(s, p) \mathbin{;} (s - p, p') \eqdf (s, p + p')$.

The small category for the nonempty lists and cyclic shifts
comonad has $S \eqdf \Nat$, $\Pt \eqdf \Sigma s : \Nat.\, [0..s]$,
$\s\, (s, p) \eqdf s$, $\t\, (s, p) \eqdf s$, $\id\, \ia{s} \eqdf (s,
0)$, $(s, p) \mathbin{;} (s, p') \eqdf (s, (p + p')
\mathbin{\mathrm{mod}} (s + 1))$.

The small category for the reader comonad has as $S$ any given set,
$\Pt \eqdf \Sigma s : S.\, 1 \cong S$, $\s\, s \eqdf s$, $\t\, s \eqdf
s$, $\id\, \ia{s} \eqdf s$, $s \mathbin{;} s \eqdf s$. This is the
discrete category on the set of objects $S$---the free category on
$S$. The small category for the array comonad has as $S$ any given
set, $\Pt \eqdf \Sigma s : S.\, S \cong S \times S$, $\s\, (s, s')
\eqdf s$, $\t\, (s, s') \eqdf s'$, $\id\, \ia{s} \eqdf (s, s)$, $(s,
s') \mathbin{;} (s', s'') \eqdf (s, s'')$. This category has a unique
morphism between any two objects and is the cofree
category on $S$.

\section{The Coproduct and a Tensor of Directed Containers}
\label{sect:coproductandtensor}

We can now look at some basic constructions of directed containers as
constructions of categories.

\subsection{Coproduct}

Given two small categories $(S_0, \Pt_0, \s_0, \t_0, \id_0, {;_0})$
and $(S_1, \Pt_1, \s_1, \t_1, \id_1, {;_1})$, we can construct a small
category $(S, \Pt, \s, \t, \id, {;})$ by 
\[
\begin{array}{c}
S \eqdf S_0 + S_1
\\
P \eqdf P_0 + P_1
\\
\s\, (\inl\, p) \eqdf \inl\, (\s_0\, p)
\\
\s\, (\inr\, p) \eqdf \inr\, (\s_1\, p)
\\
\t\, (\inl\, p) \eqdf \inl\, (\t_0\, p)
\\
\t\, (\inr\, p) \eqdf \inr\, (\t_1\, p)
\\
\id\, \ia{\inl\, s} \eqdf \inl\, (\id_0\, \ia{s})
\\
\id\, \ia{\inr\, s} \eqdf \inr\, (\id_1\, \ia{s})
\\
\inl\, p \mathbin{;} \inl\, p' \eqdf \inl\, (p \mathbin{;_0} p')
\\
\inr\, p \mathbin{;} \inr\, p' \eqdf \inr\, (p \mathbin{;_1} p')
\end{array}
\]


This small category $(S, \Pt, \s, \t, \id, {;})$ is a coproduct of 
$(S_0, \Pt_0, \s_0, \t_0, \id_0, {;_0})$
and $(S_1, \Pt_1, \s_1, \t_1, \id_1, {;_1})$ in the
category of small categories and functors, but it is likewise their
coproduct in the category of small categories and pre-opcleavages.

Via the isomorphism of the categories of directed polynomials and
directed containers, we have cheaply got a construction for the
coproduct of two directed containers. Given two directed containers
$(S_0, P_0, {\dn_0}, \o_0, {\pl_0})$ and $(S_1, P_1, {\dn_1}, \o_1,
{\pl_1})$, their coproduct is $(S, P, {\dn}, \o, {\pl})$ where 
\[
\begin{array}{c}
S \eqdf S_0 + S_1
\\
P\, (\inl\, s) \eqdf P_0\, s
\\
P\, (\inr\, s) \eqdf P_1\, s
\\
\inl\, s \dn p \eqdf \inl\, (s \mathbin{\dn_0} p)
\\
\inr\, s \dn p \eqdf \inr\, (s \mathbin{\dn_1} p)
\\
\o\, \ia{\inl\, s} \eqdf \o_0\, s
\\
\o\, \ia{\inr\, s} \eqdf \o_1\, s
\\
p \pl\, \ia{\inl\, s} p' \eqdf p \pl_0 \ia{s}\, p'
\\
p \pl\, \ia{\inr\, s} p' \eqdf p \pl_1 \ia{s}\, p'
\end{array}
\]


The interpretation $\dcsem -$ takes the coproduct of two directed containers into the
coproduct of the corresponding comonads, i.e., it preserves
coproducts.  The category of set comonads inherits its coproducts from
the category of set functors.

\subsection{A Tensor}

Given two small categories $(S_0, \Pt_0, \s_0, \t_0, \id_0, {;_0})$
and $(S_1, \Pt_1, \s_1, \t_1, \id_1, {;_1})$, we can also build a small category
$(S, \Pt, \s, \t, \id, {;})$ by 
\[
\begin{array}{c}
S \eqdf S_0 \times S_1
\\
P \eqdf P_0 \times P_1
\\
\s\, (p_0, p_1) \eqdf (\s_0\, p_0, \s_1\, p_1)
\\
\t\, (p_0, p_1) \eqdf (\t_0\, p_0, \t_1\, p_1)
\\
\id\, \ia{s_0, s_1} \eqdf (\id\, \ia{s_0}, \id\, \ia{s_1})
\\
(p_0, p_1) \mathbin{;} (p'_0, p'_1) \eqdf (p_0 \mathbin{;_0} p'_0, p_1
\mathbin{;_1} p'_1)
\end{array}
\]


In the category of small categories and functors, $(S, \Pt, \s, \t, \id, {;})$ 
is the Cartesian product of \linebreak $(S_0, \Pt_0, \s_0, \t_0, \id_0, {;_0})$ and 
$(S_1, \Pt_1, \s_1, \t_1, \id_1, {;_1})$. If we replace functors with
relative split pre-opcleavages, this ceases to be the case; the problem is that
we cannot define pairing. So we only get a binary operation on small
categories that is associative and functorial in both arguments (a
semimonoidal structure).

The corresponding construction on directed containers is the
following. Given two directed containers $(S_0, P_0, {\dn_0}, \o_0,
{\pl_0})$ and $(S_1, P_1, {\dn_1}, \o_1, {\pl_1})$, we build a new
directed container $(S, P, {\dn}, \o, {\pl})$ by 
\[
\begin{array}{c}
S \eqdf S_0 \times S_1
\\
P\, (s_0, s_1) \eqdf P\, s_0 \times P\, s_1
\\
(s_0, s_1) \dn (p_0, p_1) \eqdf (s_0 \mathbin{\dn_0} p_0, s_1 \mathbin{\dn_1} p_1)
\\
\o\, \ia{s_0, s_1} \eqdf (\o_0\, \ia{s_0}, \o_1\, \ia{s_1})
\\
(p_0, p_1) \pl (p'_0, p'_1) \eqdf (p_0 \mathbin{\pl_0} p'_0, p_1
\mathbin{\pl_1} p'_1) 
\end{array}
\]
In Glasgow, this construction on the underlying
containers has been named Hancock's tensor.


The sum and tensor of containers provide a semiring category
structure on the category of containers. The category of small
categories and split pre-opcleavages (or directed containers)
inherits this structure.





\section{The Opposite Directed Container}
\label{sect:oppositedcontainer}

Given a small category $(S, \Pt, \s, \t, \id, {;})$, an obvious
related small category to look at is the \emph{opposite} category
$(S^\op, \Pt^\op, \s^\op, \t^\op, \id^\op, {;^\op})$ where 
\[
\begin{array}{c}
S^\op \eqdf S
\\
\Pt^\op \eqdf \Pt
\\
\s^\op\, p\eqdf \t\, p
\\
\t^\op\, p \eqdf \s\, p
\\
\id^\op\, \ia{s} \eqdf \id\, \ia{s}
\\
f \mathbin{;^\op} g \eqdf g \mathbin{;} f
\end{array}
\]


Translated to directed containers, we get the following
construction. Given a directed container $(S, P, {\dn}, \o, {\pl})$,
the ``opposite'' directed container is $(S^\op, P^\op, {\dn^\op},
\o^\op, {\pl^\op})$ where 
\[
\begin{array}{c}
S^\op \eqdf S
\\
P^\op\, s \eqdf \Sigma s' : S.\, \Sigma p : P\, s'.\, \ia{s = s' \dn p}
\\
s \dn^\op (s', p) \eqdf s'
\\
\o^\op\, \ia{s} \eqdf (s, \o\, \ia{s})
\\
(s', p) \pl^\op \ia{s}\, (s'',
p') \eqdf (s'', p' \pl \ia{s''}\, p)
\end{array}
\]


For a datatype of node-labelled trees of some branching type, this
construction delivers the datatype of context-labelled trees of the
same branching type.
Let us work out what happens to the directed container for the
non-empty list and suffixes comonad.
The opposite category is given by 
\[
\begin{array}{c}
S^\op \eqdf \Nat
\\
\Pt^\op \eqdf
\Sigma s : \Nat.\, [0..s]
\\
\s^\op\, (s, p) \eqdf s - p
\\
\t^\op\,
(s, p) \eqdf \s
\\
\id^\op\, \ia{s} \eqdf (s, 0)
\\
(s - p, p')
\mathbin{;^\op} (s, p) \eqdf (s, p + p')
\end{array}
\]
%
%
Accordingly, the opposite directed container is given by 
\[
\begin{array}{c}
S^\op \eqdf
\Nat
\\
P^\op\, s \eqdf \Sigma s' : \Nat.\, \Sigma p : [0..s'].\, \ia{s =
s' - p}
\\
s \dn^\op (s', p) \eqdf s'
\\
\o^\op\, \ia{s} \eqdf (s, 0)
\\
(s', p) \pl^\op \ia{s}\, (s'', p') \eqdf (s'', p' + p)
\end{array}
\]
It is easy
to see that the second component of a position determines the first:
$s'$ must be $s + p$, so we can leave $s'$ out, removing the bound on
$p$ and letting it range over all of $\Nat$. Hence we can simplify:
\[
\begin{array}{c}
S^\op \eqdf \Nat
\\
P^\op\, s \eqdf \Nat
\\
s \dn^\op p \eqdf s + p
\\
\o^\op\, \eqdf 0
\\
p \pl^\op p' \eqdf p' + p
\end{array}
\]


We get that the underlying functor of the comonad is defined by
$D^\op\, X \eqdf \Sigma s : \Nat.\, \Nat \to X \cong \Nat \times
\Str\, X$. This is exactly the datatype of context-labelled trees of
our chosen branching type; i.e., a datatype whose every element is a
tree together with a label for every possible one-hole context it can
fill. In our example, a tree is a nonempty list over $1$, identified
with a natural number. Its contexts are longer nonempty lists. The
counit extracts the label of the empty context in a context-labelled
tree. The comultiplication replaces the label of every context with
the corresponding context-labelled tree. Formally, $\eps\, (s, xs)
\eqdf \mathsf{hd}\, xs$, $\de\, (s, xs) \eqdf (s, \de_0\, (s, xs))$
where $\de_0\, (s, xs) \eqdf (s, xs) :\!: \de_0 (s+1, \mathsf{tl}\, xs)$.

\section{Bidirected Containers as Groupoids}
\label{sect:bdcontainers}

A groupoid is a category where every morphism is iso. In algebraicized
form, a \emph{groupoid} is a category $(S, \Pt, \s, \t, \id, {;})$
together with a map ${(-)}^\inv : \Pt \to \Pt$ such that $\s\, (p^\inv)
= \t\, p$ and 
\[
\begin{array}{c}
\t\, (p^\inv) = \s\, p \\
p \mathbin{;} (p^\inv) = \id\, \ia{\s\, p} \\ 
(p^\inv) \mathbin{;} p = \id\, \ia{\t\, p}
\end{array}
\]
Here the 3rd displayed equation is welltyped because the 1st
holds. The 1st equation is in fact redundant, since it follows from
the 2nd equation together with the 1st and 2nd equations of a
category.

Translating this axiomatization to directed containers, we get what we call bidirected
containers. 

A \emph{bidirected container} is a directed container $(S, P, {\dn},
\o, {\pl})$ together with a map ${\ominus} : \Pi \ia{s : S}.\, \Pi p :
P\, s.\, P\, (s \dn p)$ such that
\[
\begin{array}{c}
(s \dn p) \dn (\ominus\, \ia{s}\, p) = s \\
p \pl \ia{s}\, (\ominus\, \ia{s}\, p) = \o\, \ia{s} \\
(\ominus\, \ia{s} p) \pl \ia{s \dn p}\, p = \o\, \ia{s \dn p}
\end{array}
\]
Again the 3rd displayed equation is welltyped because the 1st holds,
and again the 1st equation is redundant as derivable from the
2nd equation together with the 1st and 2nd equations of a directed
container.

The conversions between the two are given by $\ominus\, \ia{s}\, p \eqdf
p^\inv$ and $(s, p)^\inv \eqdf (s \dn p, \ominus\, \ia{s}\, p)$.

Intuitively, in a bidirected container, not only can positions of a
shape's subshape be translated to it, but also the other way
around. Indeed, $\ominus\, \ia{s}\, p$ should be seen as the translation of
the root position of the shape $s$ into the subshape determined by the
position $p$. 

We recall that, if a category is a groupoid, then it is isomorphic to
its dual. But the converse is not generally true. For example, the small 
category for the streams comonad is isomorphic to its own
dual (as is any category with one object), but it is not a groupoid.

The small category for the nonempty lists and cyclic shifts comonad is
a groupoid. In particular, we have $(s, p)^\inv \eqdf (s, - p
\mathbin{\mathrm{mod}} (s + 1))$. In the corresponding bidirected
container, we have $\ominus\, \ia{s}\, p \eqdf - p
\mathbin{\mathrm{mod}} (s + 1)$.

The small category for the reader comonad is also a groupoid (as a discrete
category), via $s^\inv \eqdf s$. In the corresponding bidirected
container, we have $\ominus\, \ia{s}\, s \eqdf s$.  The small category
for the array comonad is a groupoid via $(s, s')^\inv \eqdf (s',
s)$. The corresponding bidirected container has $\ominus\, \ia{s}\, s'
\eqdf s$.

\section{Conclusion}

We have witnessed that the polynomial view of containers reveals a
symmetry between shapes/subshapes of positions in the concept of a
directed container---the symmetry between sources/targets of morphisms
in the concept of a category. This makes specific constructions and
specializations available for those comonads whose underlying functors
are containers.
The concept of a directed container morphism however breaks this
symmetry; it is also quite intricate.
This situation seems to be quite special for containers that interpret
into comonads. For instance, containers that interpret into monads
do not admit a comparably simple explicit description.

As a continuation of this work, we would like to analyze our previous
results on compatible compositions and distributive laws of directed
containers \cite{AU:disldc} from the polynomial viewpoint. We expect
that this will lead to a generalization of the concepts of a Zappa-Sz\'ep
product and a matching pair of actions from monoids to small
categories.
We would also like to see if some analogs of the construction of
the opposite direct container and the concept of a bidirected
container are available for comonads directly.

\paragraph{Acknowledgements}

Ahman was funded by the Kristjan Jaak scholarship programme of the 
Archi\-medes Foundation and the Estonian Ministry of Education and
Research.  Uustalu was supported by the Estonian Ministry of Education
and Research institutional research grant no.~IUT33-13 and the
Estonian Science Foundation grant no.~9475.

\bibliographystyle{eptcs}

\newcommand{\doi}[1]{\href{http://dx.doi.org/#1}{doi: #1}}

\end{document}